\documentclass[11pt]{article}
\usepackage{jheppub}
\usepackage[utf8]{inputenc}
\usepackage{amsmath}
\usepackage{graphicx}
\usepackage{xcolor}
\usepackage{comment}
\usepackage[toc,page]{appendix}
\usepackage{feynmp}
\usepackage{tikz-feynman}
\tikzfeynmanset{compat=1.1.0}
\usetikzlibrary{positioning,arrows}
\usetikzlibrary{decorations.pathmorphing}
\usetikzlibrary{decorations.markings}

\graphicspath{{Images/}}

\title{Twist Operator BOPE and Entanglement Entropy in 2D Interface CFT}
\abstract{We address several aspects of entanglement entropy of 2D interface CFT using the replica method. Unlike the case of boundary CFT, we consider the boundary OPE (BOPE) of the R\'enyi twist operator and find a boundary twist operator anchored on the interface. This approach gives the $O(1)$ contribution to the entanglement entropy in terms of the BOPE coefficients of the twist operator. We further analyze entanglement entropy of different intervals and compare our findings with previous holographic results.}

\author{Mianqi Wang}

\affiliation{Weinberg Institute for Theoretical Physics,
University of Texas at Austin\\  2515 Speedway, Austin, TX 78712, USA.}

\emailAdd{mqwang@utexas.edu}

\begin{document}
\maketitle

\section{Introduction}
Entanglement entropy (EE) is an important quantum information quantity for two-dimensional (2D) CFT. For a spacelike interval $A=[u,v]$ with length $l$, the universal structure of EE is $S_A=\frac{c}{3}\log l/\epsilon+\text{const.}$ \cite{Pasquale_Calabrese_2004}. The coefficient of the $\log\epsilon$ term is universal, while constant terms can be altered by redefining the UV cutoff $\epsilon$.

The canonical way to derive entanglement entropy is to use replica methods to derive the partition function of the $n$-replicated theory \cite{Pasquale_Calabrese_2004,Calabrese_2009}. Concretely, one consider the density matrix of $n$ copies of the system on an $n$-sheeted manifold $\Sigma_n$ and trace out the complement region $A^c$ to get $\mathrm{Tr}(\rho_A^n)$. This is done by gluing together $A^c$ in all $n$ replica and introduce a branch cut along region $A$. The entanglement entropy of $A$ is then given by
\begin{equation}
    S_A=\lim_{n\to 1}\frac{1}{1-n}\log\mathrm{Tr}(\rho_A^n).
    \label{eq:renyi}
\end{equation} 

If we instead consider the symmetric orbifold CFT$^{\otimes n}/\mathbb{Z}_n$, we effectively squash the picture into one single plane and introduce two defects at the entangling surfaces of $A$ \cite{Pasquale_Calabrese_2004,Lashkari_2016,S_rosi_2016}. For 2D CFT, they are local twist operators, charged under the quantum $\mathbb{Z}_n$ symmetry after we gauged the $\mathbb{Z}_n$ replica symmetry. The partition function is given by its 2-pt function
\begin{equation}
    \text{Tr}(\rho_A^n)  =\frac{Z_n}{(Z_1)^n}= \epsilon^{2\Delta_n}\left\langle \sigma_n(u) \bar{\sigma}_n(v)  \right\rangle_{CFT^{\otimes n}/\mathbb{Z}_n}
    \label{eq:2dcftee}
\end{equation}

The twist operators $\sigma_n,\bar{\sigma}$ can be viewed as conformal primaries, and their dimensions are $h_n=\bar{h}_n=\frac{c}{24}(n-\frac{1}{n})$. The 2-pt function is fixed to be $1/l^{2\Delta_n}$ up to a normalization constant $C_{\sigma_n\bar{\sigma}_n}$. Notice that for clarity, we single out the UV cutoff $\epsilon$ from the normalization constant, in order to restore the dimensionless partition function. In the language of \cite{Ohmori_2015}, the piece 
$$\lim_{n\to 1}\frac{1}{1-n}\log C_{\sigma_n\bar{\sigma}_n}$$ 
is exactly the non-universal constant contributions that depends entirely on the boundary conditions at the entangling surfaces.

For 2D CFT with a conformal (Cardy) boundary, the form of EE was altered by the boundary \cite{cardy2008boundaryconformalfieldtheory}. Specifically, for $0<u<v$, the coefficient of the $\log l/\epsilon$ term is still $c/3$, while the $O(1)$ term is a non-universal function $\mathcal{F}(\eta)$ on the cross ratio $\eta=4uv/(u+v)^2$ \cite{Pasquale_Calabrese_2004}. This function depends on the full spectrum of bulk and boundary operator algebra. When $u=0$ and $A$ ends on the boundary, in the replica method, the branch cut ends naturally on the boundary, since the boundary terminates spacetime and there is no going around on the other side. In this case we only have one twist operator in our partition function, and the EE is given by its 1-pt function 
\begin{equation}
    \frac{Z_n}{(Z_1)^n}= \epsilon^{\Delta_n}\left\langle \bar{\sigma}_n(v)  \right\rangle_B=\frac{\epsilon^{\Delta_n}a_{\sigma_n}}{|2v|^{\Delta_n}}
    \label{eq:2dcftee}
\end{equation}

The result is that in addition to the universal piece $c/6\log 2l/\epsilon$, we also have a universal contribution $\log g_B$ of the Affleck-Ludwig $g$ factor to the EE that comes from the $u=0$ end. In terms of the boundary CFT (BCFT) data, 
$$\log g_B=\lim_{n\to 1}\frac{1}{1-n}\log\frac{a_{\sigma_n}}{\sqrt{C_{\sigma_n\bar{\sigma}_n}}},$$

while $a_{\sigma_n}$ and $C_{\sigma_n\bar{\sigma}_n}$ alone highly depends on the boundary conditions at the entangling surfaces \cite{Ohmori_2015}.

The main subject of this paper will be EE of interval $A$ in 2D interface CFT (ICFT), where a time-like conformal interface separates two (possibly different) CFTs. In \cite{Sakai_2008,karch2023universalityeffectivecentralcharge}, the leading $\log\epsilon$ coefficient of EE of an interval $A$ in ICFT was derived using the replica method. In \cite{Karch_2021,Karch_2023,afxonidis2025boundaryentropyfunctioninterface,afxonidis2025connectingboundaryentropyeffective}, holographic results in ICFT of order $\log\epsilon$ and $O(1)$ were verified using the Ryu-Takayanagi formula \cite{Ryu_2006}.

One may expect that EE in ICFT will not be that different from that in BCFT, as we can always fold it to become a boundary CFT with the tensor product theory CFT$_2\otimes\overline{\text{CFT}}_1$. However, results from \cite{Sakai_2008,karch2023universalityeffectivecentralcharge,Karch_2021,Karch_2023,afxonidis2025boundaryentropyfunctioninterface,afxonidis2025connectingboundaryentropyeffective} clearly distinct them: even at the level of $\log\epsilon$ term, its coefficient is changed for an interval that ends on the interface, which is related to a new universal quantity $c_\text{eff}$ in addition to the central charge.

We would like to systematically explain this distinction by studying the $O(1)$ terms using the boundary OPE (BOPE) algebra of the R\'enyi twist operator $\sigma_n$. In this approach, one explicitly sees the difference between EE of intervals ending on a conformal interface and those ending on a conformal boundary. As we will explain in detail in the next section, the difference lies in the fact that spacetime does not terminate on the interface. Consider the replica method for an interval in CFT$_1$ that ends on the interface. Even in the folded picture, the branch-cut twist operator at the entangling point on the boundary/interface is not trivial as in the case of usual BCFT, since this branch cut is in $1\otimes\overline{\text{CFT}}_1$, and operators can still go around this twist operator and have non-trivial monodromy between copies. Therefore, the BOPE channel of $\sigma_n$ fundamentally differs between BCFT and ICFT, and in particular the above identity channel on the boundary vanishes and is replaced by a genuine boundary twist operator $\hat{\sigma}_n$. This boundary twist operator is identified as the twist operator in the boundary symmetric orbifold $B^{\otimes n}/\mathbb{Z}_n$ (or interface symmetric orbifold $I^{\otimes n}/\mathbb{Z}_n$) \cite{Lunin_2002}. This boundary/interface twist operator (along with its BOPE algebra) will play the central role in deriving the EE in a general ICFT.

In past literature, CFT calculations of entanglement entropy in an interface CFT were mostly performed in the unfoliated plane after mapping the $n$-sheet to a single strip/cylinder with $n$ interfaces \cite{Sakai_2008,karch2023universalityeffectivecentralcharge}. They yield exact results for the free theories, but are too complicated for general ICFT to give any result beyond the $\log\epsilon$ coefficients. We propose that by taking a step back and studying the BCFT data of the R\'enyi twist operator of the symmetric orbifold theory, one can derive the general structure of entanglement entropy for any given interface CFT. They made the position-dependence of EE explicit unlike in previous approaches where we need to calculate the $n$-interface partition function. We also extracted meaningful quantities despite the non-universality of these $O(1)$ terms and analyzed their origins in the BOPE data. Examples are the $g$ factor \cite{PhysRevLett.67.161,Oshikawa_1997} and the interface degrees-of-freedom index $d_\text{eff}$ \cite{afxonidis2025boundaryentropyfunctioninterface,afxonidis2025connectingboundaryentropyeffective}. 

\textbf{Note added}: After version 1 of this manuscript was published, we were made aware of the Master's thesis \cite{gramaglia_2025_20071642}, which already proposed a similar twist defect BOPE approach for interface and introduced the defect twist operator $\hat{\sigma}_n$ to describe the entangling point as it approaches the interface.

\section{BOPE of twist operators}
In this section we consider an ICFT separated by a conformal interface at $x=0$. The CFT on the two sides have central charges $c_1$ and $c_2$. Consider the spacelike interval $A=[-L,r]$. We fix the left end of the interval where $L$ is large and positive acting as the IR cutoff. 

For simplicity of notations, let us set a positive number $0<d<L$. Below we will consider the following setups, as shown in Figure \ref{fig:interval}:

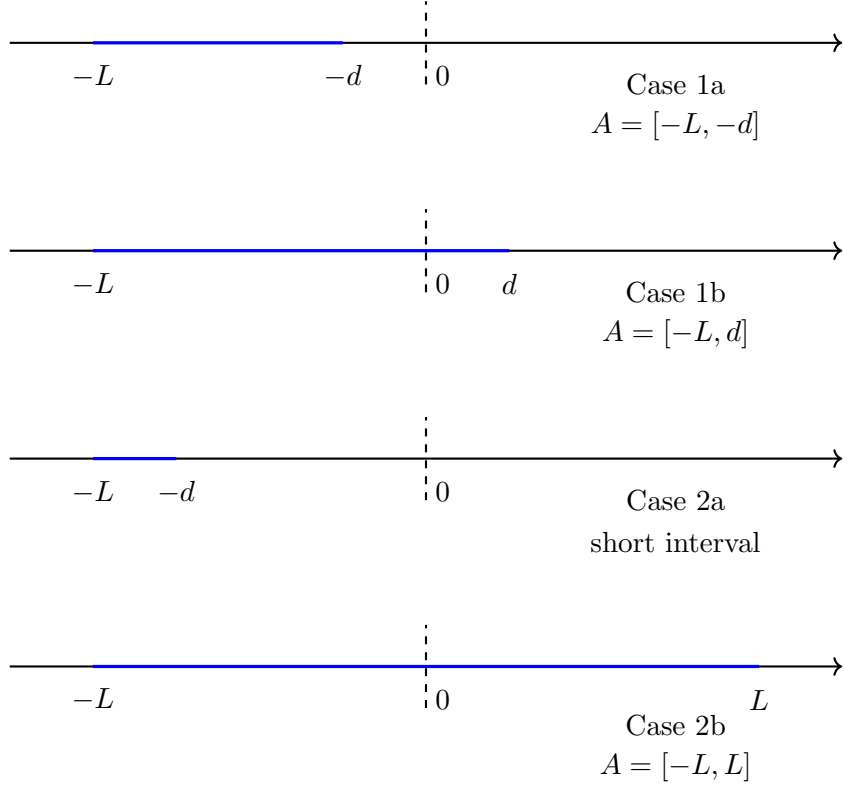
\begin{figure}[h!]
\centering
\begin{tikzpicture}[scale=1.1]

\tikzstyle{axis}=[->, thick]
\tikzstyle{interval}=[very thick, blue]
\tikzstyle{interface}=[thick, dashed]

\begin{scope}[yshift=4cm]
\draw[axis] (-5,0) -- (5,0);
\draw[interface] (0,-0.5) -- (0,0.5);

\draw[interval] (-4,0) -- (-1,0);

\node at (-4,-0.4) {$-L$};
\node at (-1,-0.4) {$-d$};
\node at (0.2,-0.4) {$0$};

\node at (3,-0.5) {Case 1a};
\node at (3,-1) {$A=[-L,-d]$};
\end{scope}

\begin{scope}[yshift=1.5cm]
\draw[axis] (-5,0) -- (5,0);
\draw[interface] (0,-0.5) -- (0,0.5);

\draw[interval] (-4,0) -- (1,0);

\node at (-4,-0.4) {$-L$};
\node at (1,-0.4) {$d$};
\node at (0.2,-0.4) {$0$};

\node at (3,-0.5) {Case 1b};
\node at (3,-1) {$A=[-L,d]$};
\end{scope}

\begin{scope}[yshift=-1cm]
\draw[axis] (-5,0) -- (5,0);
\draw[interface] (0,-0.5) -- (0,0.5);

\draw[interval] (-4,0) -- (-3,0);

\node at (-4,-0.4) {$-L$};
\node at (-3,-0.4) {$-d$};
\node at (0.2,-0.4) {$0$};

\node at (3,-0.5) {Case 2a};
\node at (3,-1) {short interval};
\end{scope}

\begin{scope}[yshift=-3.5cm]
\draw[axis] (-5,0) -- (5,0);
\draw[interface] (0,-0.5) -- (0,0.5);

\draw[interval] (-4,0) -- (4,0);

\node at (-4,-0.4) {$-L$};
\node at (4,-0.4) {$L$};
\node at (0.2,-0.4) {$0$};

\node at (3,-0.7) {Case 2b};
\node at (3,-1.2) {$A=[-L,L]$};
\end{scope}

\end{tikzpicture}

\caption{Intervals in ICFT with interface at $x=0$. Cases 1a, 1b ($L \gg d$) and 2a, 2b.}
\label{fig:interval}
\end{figure}

Case 1: $L\gg d$.

1a: $r=-d$. This is a single-sided interval entirely in CFT$_1$. In particular, it includes the case where $d=\epsilon$, which is the setup in \cite{Sakai_2008} and case (iii) in \cite{karch2023universalityeffectivecentralcharge}.

1b: $r=d$. Interval crosses the interface.

Case 2a: $r=-d$, $L\gtrsim d\gg\epsilon$ and $L-d\ll L$. Short length one-sided interval.

2b: $r=d=L$. Symmetric interval.

Case a are for one-sided intervals entirely in CFT$_1$, while case b are for intervals crossing the interface. 

Below we use the replica method to compute the entanglement entropy of the subregion $A$ to study more generally the dependence of EE on its end points in presence of the interface at $x=0$. We consider the $n$-replica theory on the complex plane $\mathbb{C}$ with twist operators inserted at the entangling points. 

In case 1, we focus on the $L\gg d$ limit and obtain the EE results for the one-sided and crossing intervals, up to $O(d/L)$. We study the difference between them and identify the quantity that describes the localized degrees of freedom of the interface. In case 2, we study the symmetric interval and the short-length interval. By subtracting the EE of the same interval without the interface we recover the $g$ factor for interface/boundary entropy. We compare them and check the results in \cite{afxonidis2025connectingboundaryentropyeffective}.
 
\subsection{Case a}

First we do case a where $r=-d$. The entanglement entropy of $A$ reads
\begin{equation}
    S_A=\lim_{n\to 1}\frac{1}{1-n}\log\epsilon^{2\Delta_n}\left\langle \sigma_n^1(-L) \bar{\sigma}_n^1(-d)  \right\rangle_{I}\,.
    \label{eq:renyiee}
\end{equation}

The index $I$ means that we calculate the 2-pt function in presence of the interface. Note that both twist operators $\sigma_n^1,\bar{\sigma}_n^1$ has dimension $h_n^1=\bar{h}_n^1=\frac{c_1}{24}(n-\frac{1}{n})$ and $\Delta_n^1=\frac{c_1}{12}(n-\frac{1}{n})$. We have again made the UV cutoff around the entangling point explicit and extracted it from the normalization of the 2-pt function.

To evaluate the above 2-pt function with an interface, we have to decompose the bulk primaries like the twist operator into defect primaries $\hat{O}$ through the interface/boundary OPE tower
\begin{equation}
    \bar{\sigma}_n^1(-d)=\sum_{\hat{O}^\dagger}\frac{b_{\bar{\sigma}_n^1\hat{O}^\dagger}}{(2d)^{\Delta_n^1-\hat{\Delta}_{\hat{O}}}}\hat{O}^\dagger(0)=\frac{a_{\bar{\sigma}_n^1}}{(2d)^{\Delta_n^1}}+\sum_{\hat{O}\neq 1}\frac{b_{\bar{\sigma}_n^1\hat{O}^\dagger}}{(2d)^{\Delta_n^1-\hat{\Delta}_{\hat{O}}}}\hat{O}^\dagger(0)
    \label{eq:bope}
\end{equation}

where we have singled out the identity channel corresponding to the 1-pt function. Here $b$ are the BOPE coefficients of the CFT$_1$ twist operator $\bar{\sigma}_n$ with the folded boundary, and we have absorbed the differential operators $D[x,\partial_t]$ into the boundary operators. We use the convention where $\hat{O}$ are normalized to 1 and $b$ absorbs all normalization for the boundary primaries. In the language of \cite{Ohmori_2015} $b$ contains all the information about the boundary conditions at the entangling surfaces that subject to change of the regulator $\epsilon$.

A very important point is that in the present case of interface (as opposed to a defect), we always treat the interface OPE in the folded picture as a BOPE \cite{Oshikawa_1997,Gliozzi_2015}: There are more CFT data than those of the two CFTs, and dimensions of operators, especially those of the interface operators, are charged under the total stress tensor $T_\text{tot}=T_2+\bar{T}_1$. Conformal primaries of CFT$_1$, such as $\sigma_n(z)$, has the same dimension $(h_n^1,\bar{h}_n^1)$ under $T_\text{tot}^{\otimes n}$. Since we are in the folded picture CFT$_2\otimes\overline{\mathrm{CFT}}_1$, $\sigma(z)$ for this interval are now in the Re$z>0$ region. The difference between case a and case b lies in whether one of the twist operators is $\bar{\sigma}_n^1$ or $\bar{\sigma}_n^2$, and their BOPE spectra. Notice that flipping CFT$_1$ causes all BOPE of $\sigma_n^1$ and $\bar{\sigma}_n^1$ to flip the signs of $\partial_x$.

However, this is not a regular BCFT situation since the interval is only in one component of the tensor product theory. The twist operator is charged under the quantum $\mathbb{Z}_n$ symmetry after gauging the replica $\mathbb{Z}_n$ symmetry, and $\sigma_n$ and $\bar{\sigma}_n$ has opposite charges. For a regular BCFT the boundary terminates spacetime and there is no going around when one bring $\sigma_n$ to the boundary. In this way the branch cut ends naturally on the boundary without the $\mathbb{Z}_n$ charge, and its one point function is nonzero.

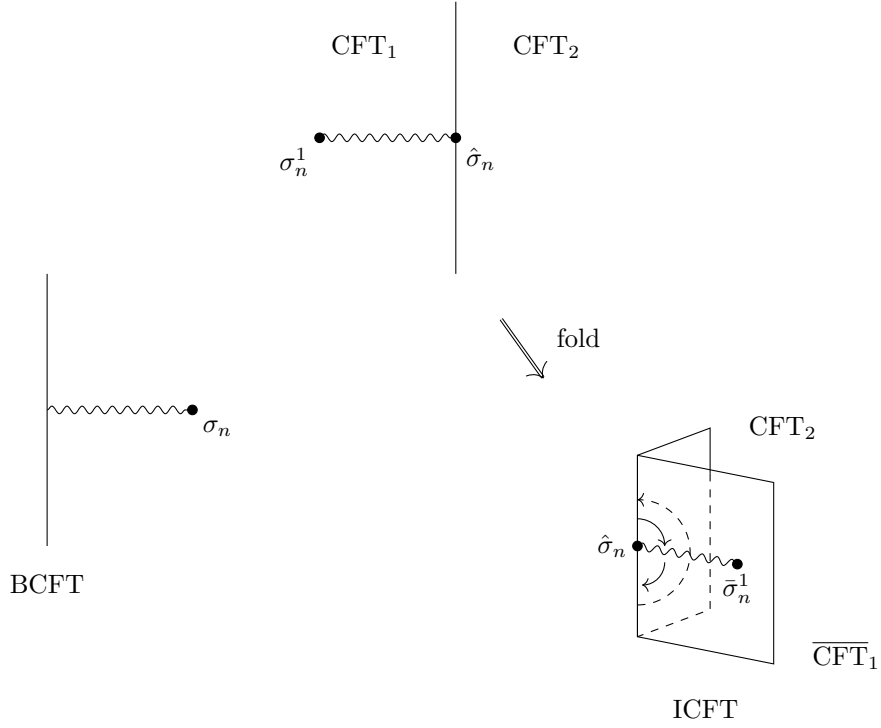
\begin{figure}[htbp]
    \centering
\begin{tikzpicture}[scale=1.2, every node/.style={font=\small}]

    \begin{scope}[shift={(0,0)}]
        \draw (0, -1.5) -- (0, 1.5);
        
        \draw[decorate, decoration={snake, segment length=2mm, amplitude=0.5mm}] (0, 0) -- (1.6, 0);
        
        \filldraw (1.6, 0) circle (1.5pt);
        \node[below right] at (1.6, 0) {$\sigma_n$};
        
        \node[below] at (0, -1.7) {BCFT};
    \end{scope}

    \begin{scope}[shift={(4.5, 3)}]
        \draw (0, -1.5) -- (0, 1.5);
        
        \node at (-1, 1) {CFT$_1$};
        \node at (1, 1) {CFT$_2$};
        
        \draw[decorate, decoration={snake, segment length=2mm, amplitude=0.5mm}] (-1.5, 0) -- (0, 0);
        
        \filldraw (0, 0) circle (1.5pt) node[below right] {$\hat{\sigma}_n$};
        \filldraw (-1.5, 0) circle (1.5pt) node[below left] {$\sigma_n^1$};
    \end{scope}

    \draw[double, ->, shorten >=2pt] (5, 1) -- (5.5, 0.3);
    \node[right] at (5.5, 0.8) {fold};

    \begin{scope}[shift={(6.5, -1.5)}]
        \draw (0, 1.0) -- (0.8, 1.3) -- (0.8, 0.8);

        \draw[dashed] (0.8,0.8) -- (0.8,-0.7) -- (0,-1);
        
        \draw (0, 1.0) -- (1.5, 0.7) -- (1.5, -1.3) -- (0, -1.0) -- cycle;
        
        \draw (0, 1.0) -- (0, -1.0);

        \node at (2.3, -1.2) {$\overline{\text{CFT}}_1$};
        \node at (1.6, 1.3) {CFT$_2$};

        \draw[->] (0, 0.3) arc (270:180:-0.3);
        \draw[->] (0.3,-0.18) arc (180:90:-0.25);
        \draw[->,dashed] (0,-0.65) arc (90:270:-0.58);

        \draw[decorate, decoration={snake, segment length=2mm, amplitude=0.5mm}] (0, 0) -- (1.1, -0.2);
        
        \filldraw (0, 0) circle (1.5pt) node[left] {$\hat{\sigma}_n$};
        \filldraw (1.1, -0.2) circle (1.5pt) node[below] {$\bar{\sigma}_n^1$};
        
        \node at (0.75, -1.8) {ICFT};
    \end{scope}

\end{tikzpicture}
\caption{The branch cut can end on the conformal boundary (left), but a boundary twist operator is required on the interface (right). Operators can still go around the boundary twist operator even in the folded picture, since the branch cut is only in $\overline{\text{CFT}}_1\otimes 1$.}

\label{fig:bope}
\end{figure}

In the present case, when we bring $\bar{\sigma}_n^1$ to the interface, operators can still go around it and reach the other copy by crossing the interface into CFT$_2$ and then crossing back. In particular, in the symmetric orbifold theory $(\text{CFT}_1\otimes_I \text{CFT}_2)^{\otimes n}/\mathbb{Z}_n$ one can have Wilson-like line operators charged under the gauged $\mathbb{Z}_n$ wrapping around the twist operator and have nontrivial commutation rules. In the folded picture, it only introduces a branch cut in $(\overline{\text{CFT}}_1\otimes 1)^{\otimes n}$, but operators gets nontrivial monodromy by covering a folded circle $S^1/\mathbb{Z}_2$ twice, with half in $\overline{\text{CFT}}_1\otimes 1$ and half in $\text{CFT}_2\otimes 1$ (See Figure \ref{fig:bope}). Thus the one point function $a_{\bar{\sigma}_n^1}$ vanishes, and the lowest boundary channel is an actual twist operator with $\mathbb{Z}_n$ charge with dimension bigger than zero. 

In addition, $\sigma_n$ and $\bar{\sigma}_n$ fuses to boundary operators $\hat{O}$ and $\hat{O}^\dagger$ that are conjugate to each other and have opposite charges under $\mathbb{Z}_n$. We also have the BOPE coefficients conjugate to each other $b_{\sigma_n^1\hat{O}}=(b_{\bar{\sigma}_n^1\hat{O}^\dagger})^*$. Then, the 2-pt function with $\sigma_n^1(L)$ is given by the bulk-to-defect correlators with the defect primaries \cite{McAvity_1995,Gliozzi_2015,Bill__2016}
\begin{equation}
    \left\langle \bar{\sigma}_n^1(L) \sigma_n^1(d)  \right\rangle_{B}=\sum_{\hat{O}\neq 1}\frac{|b_{\sigma_n^1\hat{O}}|^2}{(2d)^{\Delta_n^1-\hat{\Delta}_{\hat{O}}}(2L)^{\Delta_n^1-\hat{\Delta}_{\hat{O}}}L^{2\hat{\Delta}_{\hat{O}}}}
\end{equation}

Here notice that flipping CFT$_1$ also flips $\sigma$ to $\bar{\sigma}$. For case a, both twist operators are flipped to the right side, so we do not have $(-1)$ factors for their product $|b_{\sigma_n^1\hat{O}_*}|^2$. Since $d\le L$ we can expand the result in $d/L$ and derive the result for case a):
\begin{equation}
    \left\langle \bar{\sigma}_n^1(L) \sigma_n^1(d)  \right\rangle_{B}=\frac{|b_{\sigma_n^1\hat{O}_*}|^2}{(2d)^{\Delta_n^1-\hat{\Delta}_*}(2L)^{\Delta_n^1-\hat{\Delta}_*}L^{2\hat{\Delta}_*}}\left(1+\sum_{\hat{\Delta}_{\hat{O}}>\Delta_*}\frac{|b_{\sigma_n^1\hat{O}}|^2}{|b_{\sigma_n^1\hat{O}_*}|^2}\left(\frac{4d}{L}\right)^{\hat{\Delta}_{\hat{O}}-\Delta_*}\right)
    \label{eq:casea}
\end{equation}

In the above expression $\hat{\Delta}_*>0$ is the lowest conformal dimension above zero in the defect channel with $b_{\sigma_n^1\hat{O}_*}\neq 0$. Let us pause and analyze the BOPE tower $\hat{O}$. For an interface that does not host 1D degrees of freedom, the tower is given by boundary primaries $\partial_x^m\hat{O}_*$ and their descendants, which are $L_{-n}\dots$ acting on them. The tower is given by $\hat{\Delta}=\hat{\Delta}_*+m$ with $m\ge 0$ integers. However, once the interface has 1D degrees of freedom (DOF), the tower will be much more complicated as it contains the fusion products of $\hat{\sigma}_n$ with independent boundary primaries. From now on, DOF on the interface/boundary means that there are additional fields on the boundary in addition to the Virasoro descendants and the twist fields in the symmetric orbifold. We will discuss it more in the next section.

Let us first analyze the lowest boundary operator $\hat{O}_*$ and its dimension $\hat{\Delta}_*$. Recall that it is exactly a twist operator $\hat{O}_*\equiv\hat{\sigma}_n$ located on the interface that permutes sheets of the replica theory, since it must carry the same $\mathbb{Z}_n$ quantum number as the bulk $\sigma_n(-d)$, which carries operators $X^{(i)}$ to $X^{(i+1)}$ when they do a monodromy around $\sigma_n$. Therefore, its boundary channels must also have the same monodromy effects, thus producing twist operators. This is exactly the twist operator in the boundary symmetric orbifold $B^{\otimes n}/\mathbb{Z}_n$ in the replica method. The higher boundary operators $\hat{O}$ are either the descendants of $\hat{\sigma}_n$, or in the case that the interface hosts extra degrees of freedom, fusion of $\hat{\sigma}_n$ with other local operators.

Now let us derive its dimension. To do this we transform the $n$-sheeted BCFT picture back to the $w$-plane with $w=z^{1/n}$, and consider the correlator $\langle nT_\text{tot}(z)\hat{\sigma}_n(0)\rangle_B$. Recall that in the $w$-plane, the stress tensor is the sum of all copies: $T_\text{tot}=\sum_{i}(T_2^{(i)}+\bar{T}_1^{(i)})$. On the $w$-plane, we have a BCFT with $(n-1)$-half-line boundaries intersecting the boundary at the original point. We have
\begin{equation}
    \langle nT_\text{tot}(z)\hat{\sigma}_n(0)\rangle_B=\left\langle \left(\frac{dw}{dz}\right)^2nT_\text{tot}(w)+\frac{(c_1+c_2)n}{12}\{w;z\}\right\rangle_B\langle\hat{\sigma}_n(0)\rangle_B
\end{equation}
 
The conformal transformation in this case is similar to that in the one considered in \cite{karch2023universalityeffectivecentralcharge}, but in a folded picture. The lowest 1-pt function of $\langle T(z)\rangle_w=\left(\Delta_n^I+\frac{(c_1+c_2)}{24}\right)/z^2$ on the flat plane is nonzero in presence of the $n$-junction of boundaries/interfaces, where $\Delta_n^I$ is the lowest conformal dimension in cylinder geometry of the $n$-replicated CFT$_2\otimes_I\overline{\mathrm{CFT}}_1$. Adding in the Schwarzian terms, we get the final answer for the lowest boundary channel dimension \cite{karch2023universalityeffectivecentralcharge}
\begin{equation}
    \hat{\Delta}_*=\frac{\Delta_n^I}{n}-n\Delta_1^I 
\end{equation}

where $\Delta_1^I$ is the lowest conformal dimension of the 2-interface Hilbert space on the cylinder.

Notice that all terms have dimension $[\text{Length}^{-2\Delta_n}]$ as expected. Hence, in \eqref{eq:renyiee} we reproduce the result that in terms of expanding in $\epsilon$, the coefficient in the leading $\log\epsilon$ term of $S_A$ is as expected, $-c/3$. 

However, when $d$ is very small and comparable to the UV cutoff, i.e. $d=\epsilon\ll L$, we are left with  $\epsilon^{\Delta_n+\Delta_*}$ in \eqref{eq:renyiee} due to the presence of the $d^{\Delta_n-\hat{\Delta}_*}$ in the denominator. The coefficient of the $\log\epsilon$ term in $S_A$ is then given by\footnote{$d$ is fixed before we introduce the UV cutoff, so technically we either have $d=\epsilon\to 0$ or $d\gg\epsilon$.}
$$\lim_{n\to 1}\frac{1}{1-n}(\Delta_n+\hat{\Delta}_*)=-\frac{c+c_\text{eff}}{6}$$
where
\begin{equation}
    c_\text{eff}=\lim_{n\to 1}\frac{6}{1-n}\left(n\Delta_1^I-\frac{\Delta_n^I}{n}\right).
\end{equation}

This again recovers the results in \cite{karch2023universalityeffectivecentralcharge}. 

One can also check that for a trivial interface, \eqref{eq:bope} reduces to the ordinary Taylor expansion of $\bar{\sigma}_n^1(-d)$ at $x=0$. There $\hat{\sigma}_n=\sigma_n^1(0)$, and after folding we obtain the correct $(1-d/L)^{-2\Delta_n^1}$ for the series in the parenthesis. 

\subsection{Case b}

If $r=d$ and $A$ crosses the interface, we arrive at case b. Here we only flip $\sigma_n^1(-L)$ to the right and keep the position of $\bar{\sigma}_n^2(d)$. As mentioned in the previous subsection there will be relative signs $(-1)^m$ in the coefficients $b_{\sigma_n^1\hat{O}}$ with $\hat{\Delta}=\hat{\Delta}_*+m$. The correlator at interest is (after folding)
\begin{equation}
    \left\langle \bar{\sigma}_n^1(L) \bar{\sigma}_n^2(d)  \right\rangle_{B}=\frac{b_{\sigma_n^1\hat{\sigma}_n}^*b_{\sigma_n^2\hat{\sigma}_n}^*}{(2d)^{\Delta_n^2-\hat{\Delta}_*}(2L)^{\Delta_n^1-\hat{\Delta}_*}L^{2\hat{\Delta}_*}}\left(1+\sum_{\hat{\Delta}_{\hat{O}}>\Delta_*}\frac{b_{\sigma_n^1\hat{O}}^*b_{\sigma_n^2\hat{O}}^*}{b_{\sigma_n^1\hat{\sigma}_n}^*b_{\sigma_n^2\hat{\sigma}_n}^*}\left(\frac{4d}{L}\right)^{\hat{\Delta}_{\hat{O}}-\Delta_*}\right)
    \label{eq:caseb}
\end{equation}

Notice that the lowest boundary operator $\hat{\sigma}_n$ is the same for the BOPE of $\sigma_n^1$ and $\sigma_n^2$, since it is given by the lowest dimension in the replicated B/ICFT. However, the coefficients $b$ and higher channels might differ, as we will discuss below.

For a trivial interface, we can again check that in \eqref{eq:caseb} the correct factor $(1+d/L)^{-2\Delta_n^1}$ makes the EE only depends on $(L+d)$.

For topological interfaces, the normalization of the boundary operators $\hat{O}$ can get contribution from fusing with nontrivial fusion category elements if non-invertible symmetry is present. We then have a factor of $D^{(\frac{1}{n}-n)/2}$ on the $b$ coefficients where $D=\sum_id_i^2$ is the total quantum dimension. This gives the correct $\sqrt{D}$ contribution to the $g$ factor.

For both cases, we can then read off the $O(1)$ terms in the entanglement entropy of $A$ directly from \eqref{eq:casea} and \eqref{eq:caseb}, after taking $\log$ and $\lim_{n\to 1}$. The series summation of BOPE coefficients give exactly the effective boundary entropy $\log g^{(2)}$ defined in \cite{afxonidis2025boundaryentropyfunctioninterface,afxonidis2025connectingboundaryentropyeffective}. To derive the usual $\log g^{(1)}$ effective boundary entropy we subtract the EE with the EE without interface in the standard form $S_A=\frac{c}{3}\log \frac{l_A}{\epsilon}$.

\section{Case 1 and DOF}
Let us assume $L\gg d$ in this section and compare case 1a with case 1b. 

For case 1a where $A=[-L,-d]$, we ignore the $O(d/L)$ corrections in \eqref{eq:casea} and have
\begin{equation}
    S([-L,-d])=\frac{c_1+c_\text{eff}}{6}\log\frac{2L}{\epsilon}+\frac{c_1-c_\text{eff}}{6}\log\frac{2d}{\epsilon}+\log \tilde{g}^{(a)}+O\left(\left(\frac{d}{L}\right)^p\right).
    \label{eq:case1a}
\end{equation}

Here $\log \tilde{g}^{(a)}$ is a constant that only depends on the details of the interface (the BOPE coefficients), but not $L,d,\epsilon$:
\begin{equation}
    \log \tilde{g}^{(a)}=\lim_{n\to 1}\frac{1}{1-n}\log(2^{2\hat{\Delta}_*}|b_{\sigma_n^1\hat{O}_*}|^2)=-\frac{c_\text{eff}}{3}\log 2+\lim_{n\to 1}\frac{2\log|b_{\sigma_n^1\hat{\sigma}_n}|}{1-n}
\end{equation}

As discussed in the previous section, when the interface hosts non-trivial DOF, the gap $p>0$ between the dimension of the second lowest operator and lowest operator may not be 1 since we now have a set of nontrivial OPEs between bulk and inherent boundary primaries \cite{Pradisi_1996,Lewellen:1991tb,Pasquale_Calabrese_2004}. The set of the boundary channel is enlarged and the BOPE coefficients are very complicated. One may use techniques in \cite{Dey_2020,Kusuki_2022,Numasawa_2022} to systematically calculate the coefficients in special cases such as RCFT or Liouville.

We have shown here that the EE of the interval $A=[-L,-d]$ has dependence in the form of \eqref{eq:case1a} on its end points $L,d$ relative to the interface, up to small corrections. This is not only for the case where $A$ ends on the interface ($d=\epsilon$), but for any interval with $L\gg d$, extending the holographic results in \cite{afxonidis2025connectingboundaryentropyeffective}.

From \eqref{eq:caseb} the entanglement entropy in case 1b where $A=[-L,d]$ is

\begin{equation}
    S([-L,d])=\frac{c_1+c_\text{eff}}{6}\log\frac{2L}{\epsilon}+\frac{c_2-c_\text{eff}}{6}\log\frac{2d}{\epsilon}+\log \tilde{g}^{(b)}+O\left(\left(\frac{d}{L}\right)^p\right).
    \label{eq:case1b}
\end{equation}

Here the constant term is
\begin{equation}
    \log \tilde{g}^{(b)}=-\frac{c_\text{eff}}{3}\log 2+\lim_{n\to 1}\frac{\log (b_{\sigma_n^1\hat{\sigma}_n}^*b_{\sigma_n^2\hat{\sigma}_n}^*)}{1-n}
\end{equation}

$\log \tilde{g}^{(a)}$ and $\log \tilde{g}^{(b)}$ here are the effective boundary entropy $\log g^{(2)}$ in the notation of \cite{afxonidis2025boundaryentropyfunctioninterface}. Each individual expression for them is not meaningful since they can be changed by redefining the cutoff $\epsilon$. However, since we fix the $\epsilon$ simultaneously for both cases, we can try to extract scheme-independent quantities. 

One example is that suppose $c_1=c_2$, we have
\begin{equation}
    d_\text{eff}\equiv\log \tilde{g}^{(a)}-\log \tilde{g}^{(b)}=\lim_{n\to 1}\frac{1}{1-n}\log\left(\frac{b_{\sigma_n^1\hat{\sigma}_n}}{b_{\sigma_n^2\hat{\sigma}_n}^*}\right)
\end{equation}

If the interface does not have DOF, or if the DOF on the interface interacts with CFT$_1$ and CFT$_2$ completely symmetrically, we must have $b_{\sigma_n^1\hat{\sigma}_n}=b_{\sigma_n^2\hat{\sigma}_n}^*$ (since they have opposite $\mathbb{Z}_n$ charge and only care about the vacuum state contribution), and $d_\text{eff}$ vanishes. This applies to most of the conformal interfaces we are familiar with such as Janus \cite{Bak_2007} and free theories \cite{Oshikawa_1997,Bachas_2002}. However, in general interfaces that hosts 1D DOF, $d_\text{eff}\neq 0$ in examples such as RS braneworld \cite{Karch_2001,afxonidis2025connectingboundaryentropyeffective}. Notice that in defining this number, one do not have to demand that the interval is anchored on the interface, but only $d\ll L$. 

In addition, even when the interface hosts non-trivial DOF, the coefficient of the $\log\epsilon$ term is still universal, even though we can split the interface at least two different ways as in case 1a and 1b. This DOF is then measured by the $O(1)$ boundary entropy term. It would be interesting to study the effects of splitting interfaces with DOF in higher dimensions in the sense of \cite{Uhlemann_2023}.

\section{Case 2 and $g$ factor}

In case 2b of a symmetric interval across the interface $A=[-L,L]$, after folding, the two twist operators essentially fuse into one composite operator $\bar{\sigma}_n^1\bar{\sigma}_n^2(L)$. We can still calculate its 1pt function by using the BOPE expansion above:
\begin{equation}
    \left\langle \bar{\sigma}_n^1(L) \bar{\sigma}_n^2(L)  \right\rangle_{B}=\frac{2^{2\hat{\Delta}_*}B_n^I}{(2L)^{\Delta_n^1+\Delta_n^2}},\qquad B_n^I=\sum_{\hat{\Delta}_{\hat{O}}\ge\Delta_*}4^{\hat{\Delta}_{\hat{O}}-\Delta_*}\,b_{\sigma_n^1\hat{O}}^*b_{\sigma_n^2\hat{O}}^*
    \label{eq:sigmasigma}
\end{equation}

The entanglement entropy is
\begin{equation}
    S([-L,L])=\frac{c_1+c_2}{6}\log \frac{2L}\epsilon+\log \tilde{g}^{(2b)}+O\left(\frac{d}{L}\right)
\end{equation}

From now on suppose we have $c_1=c_2=c$. Subtracting the trivial interface case with $B_n^0=2^{-2\Delta_n}$ (under same scheme), we derived the $g$ factor of the interface, which in the folded picture is the Affleck-Ludwig $g$ factor of the boundary \cite{PhysRevLett.67.161}:
\begin{equation}
    \log g_I=\lim_{n\to 1}\frac{\log 2^{2\hat{\Delta}_*}B_n^I}{1-n}=-\frac{c_\text{eff}}{3}\log 2+\lim_{n\to 1}\frac{\log B_n^I}{1-n}
\end{equation}

Let us further assume that the interface does not host DOF, and is totally symmetric on the two sides. Then the BOPE tower is discrete and made up of $(\partial_x)^m\hat{\sigma}_n$ and their descendants. Group the boundary operators in terms of dimensions and denote $$b_m=\sum_{\hat{\Delta}=\Delta_*+m}b_{\sigma_n^1\hat{O}}=(-1)^m\sum_{\hat{\Delta}=\Delta_*+m}b_{\sigma_n^2\hat{O}}^*,$$

the $g$ factor is then
\begin{equation}
    \log g_I=-\frac{c_\text{eff}}{3}\log 2+\lim_{n\to 1}\frac{\log(\sum_{m=0}^\infty (-4)^{m}\,|b_m|^2)}{1-n}
    \label{eq:case2b}
\end{equation}

Consider case 2a where $A=[-L,-d]$ and $L,d\gg L-d$. The interval is very small, and is almost the flipped case 2b. In this case with the symmetric interface, the correlator in \eqref{eq:casea} gives, up to leading order,
\begin{equation}
    S([-L,-d])=\frac{c}{3}\log \frac{L-d}\epsilon+\log \tilde{g}^{(2a)}+O\left(\frac{L-d}{L}\right)
\end{equation}

where the effective boundary entropy $\log g^{(1)}$ (in notation of \cite{afxonidis2025connectingboundaryentropyeffective}) in this case is
\begin{equation}
    \log \tilde{g}^{(2a)}=-\frac{c_\text{eff}}{3}\log 2+\lim_{n\to 1}\frac{1}{1-n}\lim_{d\to L}\log\left[\left(1-\frac{d}{L}\right)^{2\Delta_n}\sum_{m=0}^\infty \left(\frac{4d}{L}\right)^{m}\,|b_m|^2\right]
    \label{eq:case2a}
\end{equation}

On the other hand, if we first consider the OPE between $\sigma_n^1(-L)$ and $\bar{\sigma}_n^1(-d)$, in the folded picture it is
\begin{equation}
    \left\langle \bar{\sigma}_n^1(L) \sigma_n^2(d)  \right\rangle_{B}=\sum_O\frac{\langle O\rangle_B}{(L-d)^{2\Delta_n^1-\Delta_O}}=\frac{1}{(L-d)^{2\Delta_n}}\left(1+\sum_{O>1}a_O\left(\frac{L-d}{L}\right)^{\Delta_O}\right)
\end{equation}

Here note that the identity operator is in the OPE channel of $\sigma$ and $\bar{\sigma}$. This is unlike the two-sided interval in \eqref{eq:sigmasigma} where in the folded picture it is the OPE of two charge--$(-1)$ twist operators. In that case, if one first do the OPE of $\bar{\sigma}(L)$ and $\bar{\sigma}(d)$, one do not have the identity in the OPE channel, thus producing a different result. The result here for case 2a is, to leading order, $S_A=\frac{c}{3}\log\frac{L-d}{\epsilon}$ and $\log \tilde{g}^{(2a)}=0$, same as the pure CFT result without the interface/boundary. Relating these two approaches puts a constraint on the BOPE coefficients. 

It is also obvious from the above approach that the constant term in \eqref{eq:case2b} is indeed the Affleck-Ludwig $g$ factor for the folded BCFT, which is directly given by the 1-pt function of the composite charge--$(-1)$ twist operator $\bar{\Sigma}_n(L)\equiv\bar{\sigma}_n^1\bar{\sigma}_n^2(L)$ for the tensored $(\text{CFT}_2\otimes_B\overline{\mathrm{CFT}}_1)^{\otimes n}/\mathbb{Z}_n$. This was also called the left-right one point function $a_{LR}$ for the twist operator \cite{Gliozzi_2015}. This equivalence, along with the vanishing of \eqref{eq:case2a}, can be derived using crossing symmetry of the correlators following bootstrap equations in \cite{Gliozzi_2015,Dey_2020}.

\section{Conclusions and future directions}
In this paper, we studied the entanglement entropy of a space-like interval in 2D CFT vacuum state with a time-like conformal interface. We treated the twist branch-cut operator as a conformal primary in the bulk, and considered its boundary OPE expansion. In particular, unlike EE in a single CFT with a boundary, the leading (lowest dimensional) operator on the boundary is not the identity, but a boundary twist operator in the symmetric orbifold theory of the folded interface, with dimension related to $c_\text{eff}$. Although everything is formulated in the folded BCFT, the resulting BOPE tower in case a ($A=[-L,-d]$) and case b ($A=[-L,d]$) differs because of the geometry is not terminated at the interface but folded. We then derived the entanglement entropy using the bulk-to-boundary correlator of twist operators in special cases, such as when one end of the interval is much closer to the interface than the other, or when the interval symmetric or very small and far away. 

An immediate future direction would be to use operator algebra to further understand the relation between $c_\text{eff}$, $\log g$ and $d_\text{eff}$, possibly along the lines of \cite{Cardy:1989ir,Casini_2016,afxonidis2025boundaryentropyfunctioninterface,afxonidis2025connectingboundaryentropyeffective} for $g$- and $c$- theorems, and \cite{Numasawa_2022,Kusuki_2022} for bootstrap. Bounds between $\log g$ and $d_\text{eff}$ may exist since $\log g$ gets contribution from symmetry/topology and dynamics of the interface, while $d_\text{eff}$ is about the 1D DOF localized on the interface.

It is also interesting to study the twist operator in entanglement entropy in presence of conformal defects in general dimensions \cite{Gliozzi_2015,Mezei_2015,Bianchi_2016}. In higher dimensions, the branch-cut entangling surface is a $(d-2)$-D conformal defect itself, with possible deformations of its geometry and a displacement operator $\mathcal{D}$. Its interplay with an additional conformal defect and the OPEs involving both the displacement operators of the defect and that of the entangling surface is intriguing. It may also shed light on the relation between the entanglement entropy and the energy transmission \cite{karch2024universalboundeffectivecentral}. Furthermore, studying the twist operator in the replica method using symmetric orbifold techniques and topological defect lines is a direction remains to be explored \cite{S_rosi_2016,benjamin2025generalizedsymmetriesdeformationssymmetric}.

In recent literature, defects such as the twist operator $\hat{\sigma}_n$ on the conformal interface are called a boundary condition changing (BCC) defect \cite{Choi_2024,Bhardwaj_2025}. In general, lower-dimensional defects anchoring on higher-dimensional defects are interesting to study in many contexts such as toopological defects and symTFT, but non-topological BCC defects pose even more challenges \cite{Bachas_2004,Konechny_2017}.

\section*{Acknowledgments}
The author would like to thank Andreas Karch and Hirosi Ooguri for useful discussions. This work was supported in part by DOE grant DE-SC0022021 and by a grant from the Simons Foundation (Grant 651678, AK).

\bibliographystyle{JHEP}
\bibliography{references}

@article{Bill__2016,
   title={Defects in conformal field theory},
   volume={2016},
   ISSN={1029-8479},
   url={http://dx.doi.org/10.1007/JHEP04(2016)091},
   DOI={10.1007/jhep04(2016)091},
   number={4},
   journal={Journal of High Energy Physics},
   publisher={Springer Science and Business Media LLC},
   author={Billò, Marco and Gonçalves, Vasco and Lauria, Edoardo and Meineri, Marco},
   year={2016},
   month=apr, pages={1–56} }

@misc{karch2023universalityeffectivecentralcharge,
      title={Universality of Effective Central Charge in Interface CFTs}, 
      author={Andreas Karch and Yuya Kusuki and Hirosi Ooguri and Hao-Yu Sun and Mianqi Wang},
      year={2023},
      eprint={2308.05436},
      archivePrefix={arXiv},
      primaryClass={hep-th},
      url={https://arxiv.org/abs/2308.05436}, 
}

@misc{afxonidis2025connectingboundaryentropyeffective,
      title={Connecting boundary entropy and effective central charge at holographic interfaces}, 
      author={Evangelos Afxonidis and Ignacio Carreño Bolla and Carlos Hoyos and Andreas Karch},
      year={2025},
      eprint={2507.09171},
      archivePrefix={arXiv},
      primaryClass={hep-th},
      url={https://arxiv.org/abs/2507.09171}, 
}

@misc{afxonidis2025boundaryentropyfunctioninterface,
      title={The boundary entropy function for interface conformal field theories}, 
      author={Evangelos Afxonidis and Andreas Karch and Chitraang Murdia},
      year={2025},
      eprint={2412.05381},
      archivePrefix={arXiv},
      primaryClass={hep-th},
      url={https://arxiv.org/abs/2412.05381}, 
}

@article{Karch_2021,
   title={Universal relations for holographic interfaces},
   volume={2021},
   ISSN={1029-8479},
   url={http://dx.doi.org/10.1007/JHEP09(2021)172},
   DOI={10.1007/jhep09(2021)172},
   number={9},
   journal={Journal of High Energy Physics},
   publisher={Springer Science and Business Media LLC},
   author={Karch, Andreas and Luo, Zhu-Xi and Sun, Hao-Yu},
   year={2021},
   month=sept }

@article{Karch_2023,
   title={Universal behavior of entanglement entropies in interface CFTs from general holographic spacetimes},
   volume={2023},
   ISSN={1029-8479},
   url={http://dx.doi.org/10.1007/JHEP06(2023)145},
   DOI={10.1007/jhep06(2023)145},
   number={6},
   journal={Journal of High Energy Physics},
   publisher={Springer Science and Business Media LLC},
   author={Karch, Andreas and Wang, Mianqi},
   year={2023},
   month=june }

@article{Uhlemann_2023,
   title={Splitting interfaces in 4d $ \mathcal{N} $ = 4 SYM},
   volume={2023},
   ISSN={1029-8479},
   url={http://dx.doi.org/10.1007/JHEP12(2023)053},
   DOI={10.1007/jhep12(2023)053},
   number={12},
   journal={Journal of High Energy Physics},
   publisher={Springer Science and Business Media LLC},
   author={Uhlemann, Christoph F. and Wang, Mianqi},
   year={2023},
   month=Dec }

@article{McAvity_1995,
   title={Conformal field theories near a boundary in general dimensions},
   volume={455},
   ISSN={0550-3213},
   url={http://dx.doi.org/10.1016/0550-3213(95)00476-9},
   DOI={10.1016/0550-3213(95)00476-9},
   number={3},
   journal={Nuclear Physics B},
   publisher={Elsevier BV},
   author={McAvity, D.M. and Osborn, H.},
   year={1995},
   month=Sept, pages={522–576} }

@article{Gliozzi_2015,
   title={Boundary and interface CFTs from the conformal bootstrap},
   volume={2015},
   ISSN={1029-8479},
   url={http://dx.doi.org/10.1007/JHEP05(2015)036},
   DOI={10.1007/jhep05(2015)036},
   number={5},
   journal={Journal of High Energy Physics},
   publisher={Springer Science and Business Media LLC},
   author={Gliozzi, Ferdinando and Liendo, Pedro and Meineri, Marco and Rago, Antonio},
   year={2015},
   month=May }

@article{Mezei_2015,
   title={Entanglement entropy across a deformed sphere},
   volume={91},
   ISSN={1550-2368},
   url={http://dx.doi.org/10.1103/PhysRevD.91.045038},
   DOI={10.1103/physrevd.91.045038},
   number={4},
   journal={Physical Review D},
   publisher={American Physical Society (APS)},
   author={Mezei, Márk},
   year={2015},
   month=Feb }

@article{Bianchi_2016,
   title={Rényi entropy and conformal defects},
   volume={2016},
   ISSN={1029-8479},
   url={http://dx.doi.org/10.1007/JHEP07(2016)076},
   DOI={10.1007/jhep07(2016)076},
   number={7},
   journal={Journal of High Energy Physics},
   publisher={Springer Science and Business Media LLC},
   author={Bianchi, Lorenzo and Meineri, Marco and Myers, Robert C. and Smolkin, Michael},
   year={2016},
   month=July }

@article{Ohmori_2015,
   title={Physics at the entangling surface},
   volume={2015},
   ISSN={1742-5468},
   url={http://dx.doi.org/10.1088/1742-5468/2015/04/P04010},
   DOI={10.1088/1742-5468/2015/04/p04010},
   number={4},
   journal={Journal of Statistical Mechanics: Theory and Experiment},
   publisher={IOP Publishing},
   author={Ohmori, Kantaro and Tachikawa, Yuji},
   year={2015},
   month=Apr, pages={P04010} }

@article{Calabrese_2009,
   title={Entanglement entropy and conformal field theory},
   volume={42},
   ISSN={1751-8121},
   url={http://dx.doi.org/10.1088/1751-8113/42/50/504005},
   DOI={10.1088/1751-8113/42/50/504005},
   number={50},
   journal={Journal of Physics A: Mathematical and Theoretical},
   publisher={IOP Publishing},
   author={Calabrese, Pasquale and Cardy, John},
   year={2009},
   month=Dec, pages={504005} }

@article{Pasquale_Calabrese_2004,
   title={Entanglement entropy and quantum field theory},
   volume={2004},
   ISSN={1742-5468},
   url={http://dx.doi.org/10.1088/1742-5468/2004/06/P06002},
   DOI={10.1088/1742-5468/2004/06/p06002},
   number={06},
   journal={Journal of Statistical Mechanics: Theory and Experiment},
   publisher={IOP Publishing},
   author={Pasquale Calabrese and John Cardy},
   year={2004},
   month=June, pages={P06002} }

@article{Ryu_2006,
   title={Holographic Derivation of Entanglement Entropy from the anti–de Sitter Space/Conformal Field Theory Correspondence},
   volume={96},
   ISSN={1079-7114},
   url={http://dx.doi.org/10.1103/PhysRevLett.96.181602},
   DOI={10.1103/physrevlett.96.181602},
   number={18},
   journal={Physical Review Letters},
   publisher={American Physical Society (APS)},
   author={Ryu, Shinsei and Takayanagi, Tadashi},
   year={2006},
   month=May }

@article{Numasawa_2022,
   title={Universal dynamics of heavy operators in boundary CFT2},
   volume={2022},
   ISSN={1029-8479},
   url={http://dx.doi.org/10.1007/JHEP08(2022)156},
   DOI={10.1007/jhep08(2022)156},
   number={8},
   journal={Journal of High Energy Physics},
   publisher={Springer Science and Business Media LLC},
   author={Numasawa, Tokiro and Tsiares, Ioannis},
   year={2022},
   month=Aug }

@article{Lewellen:1991tb,
    author = "Lewellen, David C.",
    title = "{Sewing constraints for conformal field theories on surfaces with boundaries}",
    reportNumber = "NSF-ITP-91-32",
    doi = "10.1016/0550-3213(92)90370-Q",
    journal = "Nucl. Phys. B",
    volume = "372",
    pages = "654--682",
    year = "1992"
}

@article{Dey_2020,
   title={Operator expansions, layer susceptibility and two-point functions in BCFT},
   volume={2020},
   ISSN={1029-8479},
   url={http://dx.doi.org/10.1007/JHEP12(2020)051},
   DOI={10.1007/jhep12(2020)051},
   number={12},
   journal={Journal of High Energy Physics},
   publisher={Springer Science and Business Media LLC},
   author={Dey, Parijat and Hansen, Tobias and Shpot, Mykola},
   year={2020},
   month=Dec }

@article{Pradisi_1996,
   title={Completeness conditions for boundary operators in 2D conformal field theory},
   volume={381},
   ISSN={0370-2693},
   url={http://dx.doi.org/10.1016/0370-2693(96)00578-3},
   DOI={10.1016/0370-2693(96)00578-3},
   number={1-3},
   journal={Physics Letters B},
   publisher={Elsevier BV},
   author={Pradisi, G and Sagnotti, A and Stanev, Ya.S},
   year={1996},
   month=July, pages={97–104} }

@article{Kusuki_2022,
   title={Analytic bootstrap in 2D boundary conformal field theory: towards braneworld holography},
   volume={2022},
   ISSN={1029-8479},
   url={http://dx.doi.org/10.1007/JHEP03(2022)161},
   DOI={10.1007/jhep03(2022)161},
   number={3},
   journal={Journal of High Energy Physics},
   publisher={Springer Science and Business Media LLC},
   author={Kusuki, Yuya},
   year={2022},
   month=Mar }

@article{PhysRevLett.67.161,
  title = {Universal noninteger ``ground-state degeneracy'' in critical quantum systems},
  author = {Affleck, Ian and Ludwig, Andreas W. W.},
  journal = {Phys. Rev. Lett.},
  volume = {67},
  issue = {2},
  pages = {161--164},
  numpages = {0},
  year = {1991},
  month = {Jul},
  publisher = {American Physical Society},
  doi = {10.1103/PhysRevLett.67.161},
  url = {https://link.aps.org/doi/10.1103/PhysRevLett.67.161}
}

@article{Cardy:1989ir,
    author = "Cardy, John L.",
    title = "{Boundary Conditions, Fusion Rules and the Verlinde Formula}",
    reportNumber = "UCSB-TH-89-06",
    doi = "10.1016/0550-3213(89)90521-X",
    journal = "Nucl. Phys. B",
    volume = "324",
    pages = "581--596",
    year = "1989"
}

@article{Casini_2016,
   title={The g-theorem and quantum information theory},
   volume={2016},
   ISSN={1029-8479},
   url={http://dx.doi.org/10.1007/JHEP10(2016)140},
   DOI={10.1007/jhep10(2016)140},
   number={10},
   journal={Journal of High Energy Physics},
   publisher={Springer Science and Business Media LLC},
   author={Casini, Horacio and Landea, Ignacio Salazar and Torroba, Gonzalo},
   year={2016},
   month=Oct }

@article{Bak_2007,
   title={Three dimensional Janus and time-dependent black holes},
   volume={2007},
   ISSN={1029-8479},
   url={http://dx.doi.org/10.1088/1126-6708/2007/02/068},
   DOI={10.1088/1126-6708/2007/02/068},
   number={02},
   journal={Journal of High Energy Physics},
   publisher={Springer Science and Business Media LLC},
   author={Bak, Dongsu and Gutperle, Michael and Hirano, Shinji},
   year={2007},
   month=Feb, pages={068–068} }

@article{Karch_2001,
   title={Open and closed string interpretation of SUSY CFT’s on branes with boundaries},
   volume={2001},
   ISSN={1029-8479},
   url={http://dx.doi.org/10.1088/1126-6708/2001/06/063},
   DOI={10.1088/1126-6708/2001/06/063},
   number={06},
   journal={Journal of High Energy Physics},
   publisher={Springer Science and Business Media LLC},
   author={Karch, Andreas and Randall, Lisa},
   year={2001},
   month=June, pages={063–063} }

@article{Choi_2024,
   title={Noninvertible Symmetry-Resolved Affleck-Ludwig-Cardy Formula and Entanglement Entropy from the Boundary Tube Algebra},
   volume={133},
   ISSN={1079-7114},
   url={http://dx.doi.org/10.1103/PhysRevLett.133.251602},
   DOI={10.1103/physrevlett.133.251602},
   number={25},
   journal={Physical Review Letters},
   publisher={American Physical Society (APS)},
   author={Choi, Yichul and Rayhaun, Brandon C. and Zheng, Yunqin},
   year={2024},
   month=Dec }

@article{Bhardwaj_2025,
   title={Generalized charges, part II: Non-invertible symmetries and the symmetry TFT},
   volume={19},
   ISSN={2542-4653},
   url={http://dx.doi.org/10.21468/SciPostPhys.19.4.098},
   DOI={10.21468/scipostphys.19.4.098},
   number={4},
   journal={SciPost Physics},
   publisher={Stichting SciPost},
   author={Bhardwaj, Lakshya and Schäfer-Nameki, Sakura},
   year={2025},
   month=Oct }

@article{Bachas_2004,
   title={Loop Operators and the Kondo Problem},
   volume={2004},
   ISSN={1029-8479},
   url={http://dx.doi.org/10.1088/1126-6708/2004/11/065},
   DOI={10.1088/1126-6708/2004/11/065},
   number={11},
   journal={Journal of High Energy Physics},
   publisher={Springer Science and Business Media LLC},
   author={Bachas, Constantin and Gaberdiel, Matthias},
   year={2004},
   month=Nov, pages={065–065} }

@article{Konechny_2017,
   title={RG boundaries and interfaces in Ising field theory},
   volume={50},
   ISSN={1751-8121},
   url={http://dx.doi.org/10.1088/1751-8121/aa60f6},
   DOI={10.1088/1751-8121/aa60f6},
   number={14},
   journal={Journal of Physics A: Mathematical and Theoretical},
   publisher={IOP Publishing},
   author={Konechny, Anatoly},
   year={2017},
   month=Mar, pages={145403} }

@article{S_rosi_2016,
   title={Relative entropy of excited states in two dimensional conformal field theories},
   volume={2016},
   ISSN={1029-8479},
   url={http://dx.doi.org/10.1007/JHEP07(2016)114},
   DOI={10.1007/jhep07(2016)114},
   number={7},
   journal={Journal of High Energy Physics},
   publisher={Springer Science and Business Media LLC},
   author={Sárosi, Gábor and Ugajin, Tomonori},
   year={2016},
   month=July }

@article{Lashkari_2016,
   title={Modular Hamiltonian for Excited States in Conformal Field Theory},
   volume={117},
   ISSN={1079-7114},
   url={http://dx.doi.org/10.1103/PhysRevLett.117.041601},
   DOI={10.1103/physrevlett.117.041601},
   number={4},
   journal={Physical Review Letters},
   publisher={American Physical Society (APS)},
   author={Lashkari, Nima},
   year={2016},
   month=July }

@misc{karch2024universalboundeffectivecentral,
      title={Universal Bound on Effective Central Charge and Its Saturation}, 
      author={Andreas Karch and Yuya Kusuki and Hirosi Ooguri and Hao-Yu Sun and Mianqi Wang},
      year={2024},
      eprint={2404.01515},
      archivePrefix={arXiv},
      primaryClass={hep-th},
      url={https://arxiv.org/abs/2404.01515}, 
}

@misc{benjamin2025generalizedsymmetriesdeformationssymmetric,
      title={Generalized Symmetries and Deformations of Symmetric Product Orbifolds}, 
      author={Nathan Benjamin and Suzanne Bintanja and Yu-Jui Chen and Michael Gutperle and Conghuan Luo and Dikshant Rathore},
      year={2025},
      eprint={2509.12180},
      archivePrefix={arXiv},
      primaryClass={hep-th},
      url={https://arxiv.org/abs/2509.12180}, 
}

@misc{cardy2008boundaryconformalfieldtheory,
      title={Boundary Conformal Field Theory}, 
      author={John Cardy},
      year={2008},
      eprint={hep-th/0411189},
      archivePrefix={arXiv},
      primaryClass={hep-th},
      url={https://arxiv.org/abs/hep-th/0411189}, 
}

@article{Bachas_2002,
   title={Permeable conformal walls and holography},
   volume={2002},
   ISSN={1029-8479},
   url={http://dx.doi.org/10.1088/1126-6708/2002/06/027},
   DOI={10.1088/1126-6708/2002/06/027},
   number={06},
   journal={Journal of High Energy Physics},
   publisher={Springer Science and Business Media LLC},
   author={Bachas, Constantin and Boer, Jan de and Dijkgraaf, Robbert and Ooguri, Hirosi},
   year={2002},
   month=June, pages={027–027} }

@article{Oshikawa_1997,
   title={Boundary conformal field theory approach to the critical two-dimensional Ising model with a defect line},
   volume={495},
   ISSN={0550-3213},
   url={http://dx.doi.org/10.1016/S0550-3213(97)00219-8},
   DOI={10.1016/s0550-3213(97)00219-8},
   number={3},
   journal={Nuclear Physics B},
   publisher={Elsevier BV},
   author={Oshikawa, Masaki and Affleck, Ian},
   year={1997},
   month=June, pages={533–582} }

@article{Sakai_2008,
   title={Entanglement through conformal interfaces},
   volume={2008},
   ISSN={1029-8479},
   url={http://dx.doi.org/10.1088/1126-6708/2008/12/001},
   DOI={10.1088/1126-6708/2008/12/001},
   number={12},
   journal={Journal of High Energy Physics},
   publisher={Springer Science and Business Media LLC},
   author={Sakai, Kazuhiro and Satoh, Yuji},
   year={2008},
   month=Dec, pages={001–001} }

@article{Lunin_2002,
   title={Three-Point Functions for $M^N / S_N$ Orbifolds with $mathcal{N}=4 $ Supersymmetry},
   volume={227},
   ISSN={1432-0916},
   url={http://dx.doi.org/10.1007/s002200200638},
   DOI={10.1007/s002200200638},
   number={2},
   journal={Communications in Mathematical Physics},
   publisher={Springer Science and Business Media LLC},
   author={Lunin, Oleg and Mathur, Samir D.},
   year={2002},
   month=May, pages={385–419} }

@misc{gramaglia_2025_20071642,
  author       = {Gramaglia, Elisa},
  title        = {Energy and entropy in critical systems with
                   impurities
                  },
  month        = apr,
  year         = 2025,
  publisher    = {Zenodo},
  doi          = {https://doi.org/10.5281/zenodo.20071642},
  url          = {https://doi.org/10.5281/zenodo.20071642},
}
\end{document}